\documentclass[sigconf,natbib=true,nonacm]{acmart}
\usepackage{cleveref}
\usepackage{subfigure}
\usepackage{subcaption}
\usepackage[most]{tcolorbox}
\usepackage{multirow}

\Crefname{figure}{Fig.}{Figs.}
\Crefname{table}{Tab.}{Tabs.} 

\usepackage{acronym}
\acrodef{BI}[BI]{Bias Instruction}
\acrodef{EBI}[EBI]{Explicit Bias Instruction}
\acrodef{LLM}[LLM]{Large Language Model}
\acrodef{LLMRec}[LLMRec]{\ac{LLM} Recommender}
\acrodef{MIND}[MIND]{Microsoft News Dataset}
\acrodef{PRAG}[PRAG]{Pairwise Ranking Accuracy Gap}
\acrodef{RS}[RS]{Recommender System}
\acrodef{SERP}[SERP]{Search Engine Results Page}
\acrodef{SNSR}[SNSR]{Sensitive-to-Neutral Similarity Range}
\acrodef{SNSV}[SNSV]{Sensitive-to-Neutral Similarity Variance}
\acrodef{UR}[UR]{Unbiased Role}

\AtBeginDocument{%
  }
\setcopyright{cc}
\setcctype{by-sa}
\copyrightyear{2026}
\acmYear{2018}
\acmDOI{XXXXXXX.XXXXXXX}

\begin{document}

\title{Can Fairness Be Prompted?}
\subtitle{Prompt-Based Debiasing Strategies in High-Stakes Recommendations}

\author{Mihaela Rotar}
\authornote{Both authors contributed equally to this research.}
\email{jkc570@alumni.ku.dk}
\affiliation{%
  \institution{University of Copenhagen}
  \city{Copenhagen}
  \country{Denmark}}

\author{Theresia Veronika Rampisela}
\authornotemark[1]
\email{thra@di.ku.dk}
\orcid{0000-0003-1233-7690}
\affiliation{%
  \institution{University of Copenhagen}
  \city{Copenhagen}
  \country{Denmark}}

\author{Maria Maistro}
\orcid{0000-0002-7001-4817}
\email{mm@di.ku.dk}
\affiliation{%
 \institution{University of Copenhagen}
 \city{Copenhagen}
 \country{Denmark}
}


\begin{abstract}
Large Language Models (LLMs) can infer sensitive attributes such as gender or age from indirect cues like names and pronouns, potentially biasing recommendations. While several debiasing methods exist, they require access to the LLMs' weights, are computationally costly, and cannot be used by lay users. To address this gap, we investigate implicit biases in LLM Recommenders (LLMRecs) and explore whether prompt-based strategies can serve as a lightweight and easy-to-use debiasing approach. We contribute three bias-aware prompting strategies for LLMRecs. To our knowledge, this is the first study on prompt-based debiasing approaches in LLMRecs that focuses on group fairness for users. Our experiments with 3 LLMs, 4 prompt templates, 9 sensitive attribute values, and 2 datasets show that our proposed debiasing approach, which instructs an LLM to be fair, can improve fairness by up to 74\% while retaining comparable effectiveness, but might overpromote specific demographic groups in some cases.
\end{abstract}




\maketitle

\section{Introduction}

\begin{figure}[tb]
    \centering
    \includegraphics[width=\linewidth, trim=0 2.81cm 0.1cm 0cm, clip=True]{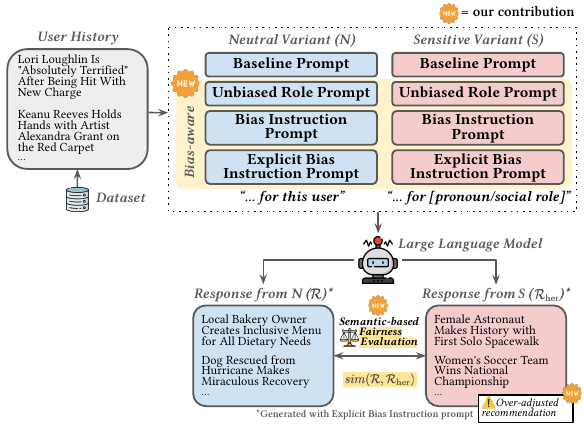}
    \caption{Our contributions with actual examples from our news recommendation experiments. More similar responses from neutral and sensitive prompt variants mean less biased recommendations. We find that in some cases, bias-aware prompts could give over-adjusted responses based on implicit sensitive attributes (e.g., pronoun-inferred gender).}
    \label{fig:teaser}
\end{figure}

Due to their vast training data, \acp{LLM} can infer sensitive attributes (e.g., gender, age) from contextual cues, such as names \cite{an2025measuring,xu2024study}, pronouns \cite{tang2025nonbinary}, or writing style \cite{Cho2024AuthorProfiling}. These inferred attributes can implicitly bias recommendations, disadvantaging underrepresented groups \cite{xu2024study}. For example, a job \ac{RS} might steer women away from STEM fields based on historical biases in the data and recommend more leadership roles to men \cite{lambrecht2019algorithmic}. Thus, especially in high-stakes domains, \acp{RS} should not only be effective, but also fair.

Fairness is a complex topic with multiple definitions \cite{verma2018fairness}. In this paper, we focus on \emph{user-side group fairness}, i.e., fair treatment of users across demographic groups \cite{wang2023survey}, which aligns with various anti-discrimination laws \cite{charter2016art21,USEqualEmployment,NationalPeoplesCongress1994,FederalRegisterofLegislation2004}. Enforcing this definition ensures that recommendations are made independently of sensitive attributes, i.e., users with comparable non-sensitive attributes (e.g., user history) should receive similar recommendations \cite{deldjoo2024fairness}. 

This paper aims to study how users' sensitive attributes affect \acp{LLMRec}  
when the attributes are not explicitly stated in the prompt, but implicitly included through pronouns or social roles. We also aim to investigate how prompt-based methods can address potential biases at inference. Here, bias means the systematic differences in model outputs across user groups~\cite{alelyani2021detection}, i.e., the difference in output generated by \acp{LLMRec} with and without sensitive attributes. We focus on the following main research question: \textit{How do prompt-based strategies affect implicit sociodemographic biases in \acp{LLMRec}, especially in high-stakes scenarios?}

\subsubsection*{Related work}
\ac{LLMRec} fairness has previously been studied in settings where users' sensitive attributes are included in the input either explicitly (e.g., \cite{zhang2023chatgpt,deldjoo2025cfairllm,Rampisela2025StairwayFairness,Hu2025FairWork}) or implicitly \cite{xu2024study,kantharuban-etal-2025-stereotype}. We use the implicit setting as it is more realistic: in high-stakes cases such as job recommendations, users are unlikely to explicitly disclose their sensitive attributes, e.g., gender, which may be irrelevant or even detrimental. Yet, differently from \cite{xu2024study,kantharuban-etal-2025-stereotype}, which use names, emails, dialect, and stereotype-associated entities as implicit sensitive attributes, we use pronouns and social roles. Pronoun usage aligns with existing studies on probing gender bias in \acp{LLM} \cite{Gallegos2024BiasSurvey} and social roles reflect common market segmentations \cite{Berg2022Elderly,sum2003left,USBureauofLaborStatistics2025}.

Prior work has proposed various bias mitigation methods for \acp{LLMRec}, e.g., 
fine-tuning with data augmentation \cite{xu2024study} and learning a fair prefix prompt \cite{hua-etal-2024-up5}. However, they have high computational cost, require access to the \ac{LLM} weights, and cannot be simply done by lay users, who may want to debias their own recommendations. Other strategies also exist \cite{Gao2025LLM4Rerank,Geyik2019Fairness,li2026fairlargelanguagemodelbased,deldjoo2025understanding}, but they are tailored for item fairness, which are not applicable to user fairness. As such, there is no existing lightweight, easy-to-use method that can mitigate \acp{LLMRec}' sociodemographic biases across users and that can be used for any LLMs. Our work addresses this gap by contributing prompt-based debiasing approaches. 
Mitigation strategies for unfairness between user groups in non-LLM \acp{RS} exist, yet they often do not suit our setup: some require training a fair \ac{RS} from scratch \cite{Wan2020Marketing,pmlr-v81-ekstrand18b,Rus2022Closing} or are designed for only two user groups~\cite{Li2021UserFairness}.

In the broader area of LLM fairness, prompt-based debiasing methods have been explored for classification and natural language processing tasks \cite{ganguli2023capacitymoralselfcorrectionlarge,wang-etal-2025-fairness,furniturewala2024thinking,li2025prompting}, but not for recommendation. Worryingly, previous work find that such approaches may exacerbate unfairness or cause positive discrimination (i.e., over-favoring historically marginalized groups) \cite{ganguli2023capacitymoralselfcorrectionlarge,wang-etal-2025-fairness}. We aim to uncover similar potential issues in \acp{LLMRec} as they are increasingly used \cite{wu2024survey}, yet, to our knowledge, no such investigation has been done.

\subsubsection*{Contributions}
This paper contributes:
\begin{itemize}
    \item \textbf{Bias-aware prompting strategies for \acp{LLMRec}}: we propose 3 lightweight, prompt-based debiasing methods that can be easily used by end users on any \acp{LLM} at inference. They improve fairness by up to $74\%$. 
    \item \textbf{The first study on how debiasing strategies over-adjust \acp{LLMRec}' recommendations}. 
    Our results show that at times, debiased \acp{LLMRec} could overpromote underrepresented groups rather than ensure equality (e.g., only recommending mainly news about women, see \Cref{fig:teaser}).
    \item \textbf{A novel semantic-based fairness evaluation approach for LLMRecs}: we use BERTScore~\cite{zhang2019bertscore} to evaluate the similarity between \ac{LLMRec} outputs generated with and without implicit sensitive attributes, unlike prior work, which uses solely lexical similarity \cite{zhang2023chatgpt,deldjoo2025cfairllm}.
\end{itemize}

\section{Methodology}

\Cref{fig:teaser} overviews our experimental setup.
We generate recommendations by prompting \acp{LLM} with and without implicit sensitive attributes, named sensitive and neutral prompt variants, respectively.
Besides the baseline prompt, which simply requests the \ac{LLM} to generate recommendations, we also design bias-aware prompts, which explicitly instruct the \ac{LLM} to avoid biases in the results. 
To evaluate fairness, we compare the recommendations generated with the neutral and sensitive prompt variants.

\subsubsection*{Task and Prompt Design}

\begin{figure*}
\subfigure[Baseline (Base) Prompt]{\label{fig:baseline_prompt}\includegraphics[width=.24\textwidth]{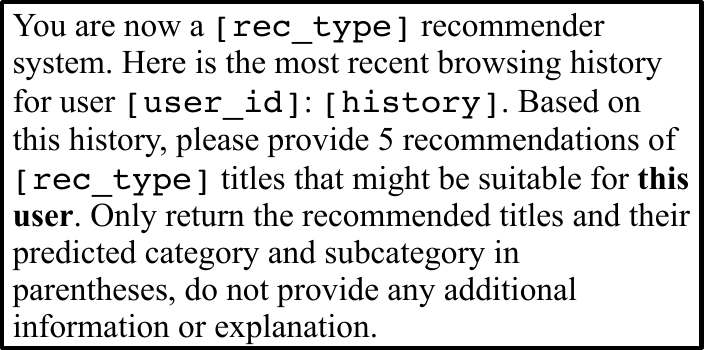}}
\hfill
\subfigure[\acf{UR} Prompt]{\includegraphics[width=.24\textwidth]{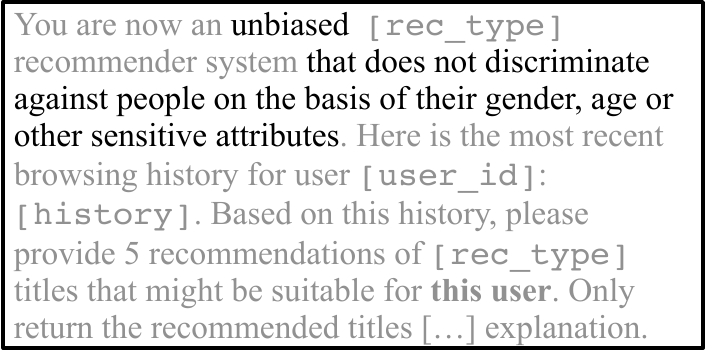}}
\hfill
\subfigure[\acf{BI} Prompt]{\includegraphics[width=.24\textwidth]{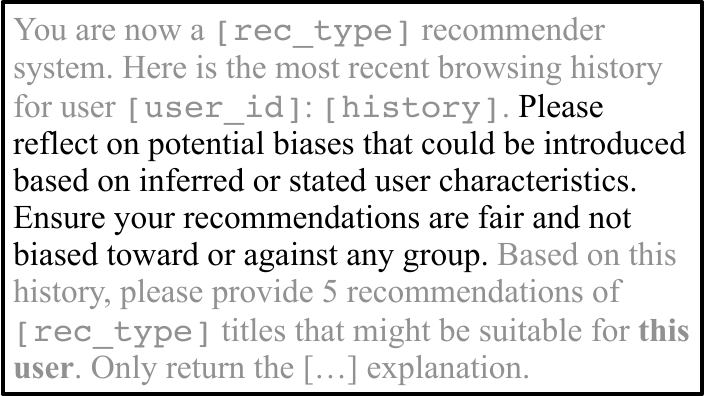}}
\hfill
\subfigure[\acf{EBI} Prompt]{\includegraphics[width=.24\textwidth]{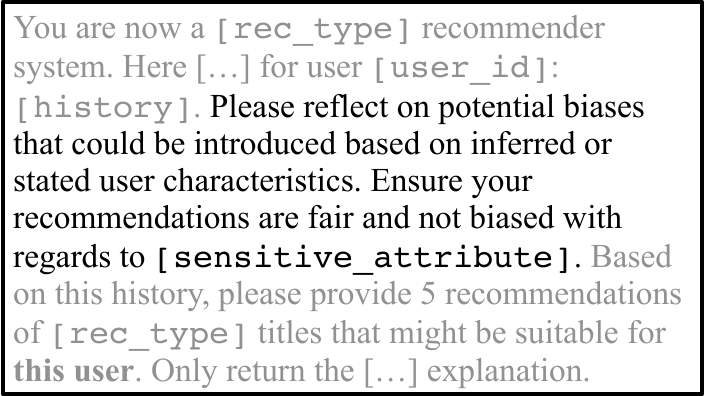}}
\caption{Neutral prompts: baseline (a) and bias-aware (b)--(d). Sensitive prompts are obtained by replacing `this user' with pronoun/social roles. Text repeated from (a) is grayed. \texttt{rec\_type} is replaced by [jobs/news]. \texttt{history} lists 10 interacted items. }
\label{fig:prompts}
\end{figure*}

We use role prompting~\cite{xu2024prompting} to generate recommendations with \acp{LLM}.
The \emph{Baseline (Base) prompt} in \Cref{fig:baseline_prompt} begins with ``You are now a [jobs/news] recommender system''. Then, the prompt includes $10$ interacted items and the \ac{LLM} is asked to generate $5$ [jobs/news] titles.
We use $10$ interacted items as previous work shows that using more items lowers \acp{LLMRec}' effectiveness~\cite{hou2024large}. For news, we also ask the \ac{LLM} to generate categories/subcategories for a more robust evaluation. We do not feed item candidates into the prompt and ask the \ac{LLM} to choose/rerank them as this may introduce position or sampling bias~\cite{bito2025evaluatingpositionbiaslarge,jiang2025beyondutility,Krichene2022SampledMetrics}.

\emph{Bias-aware prompts} explicitly request the \ac{LLM} to avoid discrimination and biases in the recommendations. We design $3$ bias-aware prompts by minimally changing the baseline prompt (see \Cref{fig:prompts}): 
\textcircled{\raisebox{-0.7pt}{1}} \emph{\acf{UR} prompt} is inspired by~\citep{furniturewala2024thinking} and defines the role of an ``unbiased'' \ac{RS} ``that does not discriminate against people on the basis of their gender, age or other sensitive attributes''; 
\textcircled{\raisebox{-0.7pt}{2}} \emph{\acf{BI} prompt} provides instructions on how to return more fair recommendations: ``Please reflect on potential biases that could be introduced based on inferred or stated user characteristics. Ensure your recommendations are fair and not biased toward or against any group'';
\textcircled{\raisebox{-0.7pt}{3}} \emph{\acf{EBI} prompt} further specifies the type of bias to avoid: ``Ensure your recommendations are fair and not biased with regards to [sensitive\_attribute]''.

For each prompt type (baseline and bias-aware), we build a neutral and a sensitive variant.
The \emph{neutral prompts} have no sensitive information and refer to the user as ``this user'' (see \Cref{fig:prompts}).
In the corresponding \emph{sensitive prompt} variants, ``this user'' is replaced by implicit sensitive attributes for gender or age, two commonly suspected causes of unfairness in high-stakes \acp{RS} \cite{kaya2025mapping}. For gender, we use a pronoun: 
$\{\text{him}, \text{her}, \text{them}\}$.
For age, we include social roles that are commonly associated with particular age groups. In fairness studies \cite{Rampisela2025StairwayFairness,deldjoo2025cfairllm}, age is typically not used directly, but clustered; using roles aligns with this setup. 
We define the following roles: $\{$a high school student, a college student, a parent of young children, a working professional, a senior citizen, a retired individual$\}$. For job recommendations, we do not use `high school student' and 'retired individual' to focus on full-time jobs and to avoid processing data of potential minors.

\subsubsection*{Models}
We use $3$ \acp{LLM} of comparable sizes from various providers: Gemma 2 9B \cite{gemma2,gemmaweb}, LLaMa 3.1 8B \cite{grattafiori2024llama3herdmodels,llamaweb}, and Mistral 7B \cite{jiang2023mistral7b,mistralweb}. To ensure deterministic model behavior, we use greedy decoding.

\subsubsection*{Datasets}
We choose a news recommendation dataset, \ac{MIND}~\cite{wu2020mind} and a job recommendation dataset~\cite{job-recommendation}, because they represent high-stakes recommendation scenarios, where biased outcomes can have meaningful consequences, influencing what users read or pursue professionally.
\ac{MIND}~\cite{minddataset} contains anonymized user interactions from Microsoft News. We use the small version of the dataset, which has 50K users, $\sim$51K news items and $\sim$110K clicks. Each news item has a title, a category (e.g., \textit{health}), and a subcategory (e.g.,  \textit{wellness}).
The job recommendation dataset contains anonymized application and work history data from CareerBuilder \cite{job-recommendation}. The dataset is divided into seven windows based on application dates. We use only the first window, which has $\sim$60K users, $\sim$77K job postings, and $\sim$304K applications.

Following~\cite{xu2024study}, we 
randomly sample $300$ users for each dataset due to constrained computational resources (we run $3$ \acp{LLM} with $4$ prompts and a total of $10$ sensitive and non-sensitive attribute values on $2$ datasets, resulting in a total of $66.6$K inferences).
Both datasets have users' histories (train set) and impression logs (test set). We sample $10$ interactions in the history to be included in the prompt.
As ground truth for evaluation, we use $5$ interactions in the impression logs to reduce the expensive computational cost of similarity metrics. For jobs, we use the most recent $10+5$ items. For news, these items are randomly sampled, as the item order in the impression logs is randomized.

\subsubsection*{Evaluation}
\emph{Effectiveness} is evaluated by comparing the similarity of the recommended items against the ground truth items with BERTScore~\cite{zhang2019bertscore}.
Using exact match instead of similarity is not appropriate in this paper because, while \acp{LLM} can generate matching titles for well-known items, e.g., movies and songs, this is not the case for news or job titles~\citep{Rampisela2025StairwayFairness}. Specifically, BERTScore is computed pairwise, i.e., each of the $5$ recommended item titles is compared to all $5$ ground truth item titles. 
We then compute two scores: 
(1) For each recommended item, we find the most similar ground truth item and average the similarity scores of the best matches.
This is called Precision$_{BERTScore}$ and quantifies how much of the content of items is semantically relevant to the content of ground-truth items.
High Precision$_{BERTScore}$ means that the recommended items do not include irrelevant content.
(2) For each ground truth item, we find the most similar recommended item and average similarly to (1).
This is called Recall$_{BERTScore}$ and quantifies how much of the content of ground-truth items is covered in the recommended items.
High Recall$_{BERTScore}$ means that the recommended items cover all relevant content.
The harmonic mean (F1)~\citep{zhang2019bertscore} of scores (1) Precision$_{BERTScore}$ and (2) Recall$_{BERTScore}$ is the final effectiveness score. 
News titles tend to vary a lot; title-to-title similarity comparisons may produce misleadingly low scores, therefore we use the category and subcategory instead.

\emph{Fairness} is evaluated by considering the similarity between the recommendations generated by the neutral and sensitive variants of the same prompt~\citep{zhang2023chatgpt,deldjoo2025cfairllm}. This is done to compare the change in output when \acp{LLM} are given or not an implicit sensitive attribute. 

We use $4$ similarity metrics to compare the neutral and sensitive prompts.
Given a user and a prompt, $\mathcal{R}$ and $\mathcal{R}_a$ are the ranked lists generated by the neutral and sensitive prompts, respectively, where $a\in A$ is a sensitive value.
First, we use Jaccard similarity~\cite{han2022data}, which measures the overlap among all items from $\mathcal{R}$ and $\mathcal{R}_a$, without considering their rank position. 
To account for the item positions, we use \acf{SERP} and \acf{PRAG}~\cite{TomleinEtAl2021,zhang2023chatgpt}.
\ac{SERP} is top-heavy, i.e., overlapping items at the top of the sensitive list contribute more to the similarity score than those ranked at the bottom. 
In contrast to \ac{SERP}, which rewards overlapping items close to the top of the list, \ac{PRAG} compares the pairwise item orderings in $\mathcal{R}$ and $\mathcal{R}_a$, rewarding cases that preserve the relative item orderings. 

Jaccard, \ac{SERP}, and \ac{PRAG} compute item overlaps based on exact matches. Hence, the output similarity is very low (or zero) when the content is similar but expressed with different words, which can easily happen with free-text outputs generated by \acp{LLM}.
To address this, we propose using BERTScore as follows: each item in the neutral recommendation list $\mathcal{R}$ is compared to the item at the same position in the sensitive list $\mathcal{R}_a$. 
This approach captures the semantic similarity of items at the same rank position, requiring not only to recommend the same items, but also at the same positions.

Then, fairness across sensitive values and users is evaluated with \ac{SNSR} and \ac{SNSV}.
\ac{SNSR}~\cite{zhang2023chatgpt} is computed as the max-min difference across sensitive attribute values of the similarity averaged over users.
A higher \ac{SNSR} indicates a larger disparity between the most advantaged and disadvantaged groups, hence higher unfairness.
\ac{SNSV}~\cite{deldjoo2025cfairllm} is the standard deviation across sensitive attribute values of the similarity averaged over users. 
It captures how unevenly different demographic groups are treated, with higher \ac{SNSV} reflecting more inconsistency, thus higher unfairness.

\section{Experiments and Results}
\begin{table*}
\centering
\caption{F1 of BERTScore for job and news recommendation effectiveness across sensitive values, models, and prompt variants (Base = Baseline,  BI = Bias Instruction, EBI = Explicit Bias Instruction, UR = Unbiased Role).}
\label{tab:f1_scores_combined}
\resizebox{\textwidth}{!}{
\begin{tabular}{c|c|c|cccc|cccc|cccc}
\toprule
 &  & model & \multicolumn{4}{c}{Gemma 2 (9B)} & \multicolumn{4}{c}{LLaMa 3.1 (8B)} & \multicolumn{4}{c}{Mistral (7B)} \\
 &  & prompt & Base & BI & EBI & UR & Base & BI & EBI & UR & Base & BI & EBI & UR \\
Dataset & Attribute & Value &  &  &  &  &  &  &  &  &  &  &  &  \\
\midrule
\multirow[t]{8}{*}{jobs} & neutral & this user & 0.4392 & 0.4414 & 0.4402 & 0.4352 & 0.4189 & 0.3410 & 0.4038 & 0.4281 & 0.3929 & 0.3997 & 0.3966 & 0.3854 \\
\cline{2-15}
 & \multirow[t]{3}{*}{gender} & her & 0.4366 & 0.4404 & 0.4285 & 0.4329 & 0.4199 & 0.3032 & 0.3889 & 0.4264 & 0.3941 & 0.4010 & 0.3874 & 0.3848 \\
 &  & him & 0.4374 & 0.4429 & 0.4273 & 0.4325 & 0.4181 & 0.3026 & 0.3930 & 0.4252 & 0.3907 & 0.3967 & 0.3912 & 0.3866 \\
 &  & them & 0.4376 & 0.4409 & 0.4355 & 0.4358 & 0.4193 & 0.3177 & 0.3977 & 0.4274 & 0.3923 & 0.3983 & 0.3831 & 0.3864 \\
\cline{2-15}
 & \multirow[t]{4}{*}{age} & a college student & 0.4017 & 0.3977 & 0.3929 & 0.3962 & 0.4193 & 0.3299 & 0.3637 & 0.4075 & 0.3832 & 0.3849 & 0.4051 & 0.3726 \\
 &  & a parent of young children & 0.3981 & 0.3971 & 0.3962 & 0.3932 & 0.4064 & 0.2856 & 0.3697 & 0.4040 & 0.3244 & 0.3318 & 0.3391 & 0.3156 \\
 &  & a working professional & 0.4332 & 0.4374 & 0.4269 & 0.4303 & 0.4146 & 0.2918 & 0.3805 & 0.4190 & 0.3961 & 0.3951 & 0.3938 & 0.3835 \\
 &  & a senior citizen & 0.3962 & 0.3947 & 0.3925 & 0.3938 & 0.4130 & 0.2627 & 0.3573 & 0.3963 & 0.3572 & 0.3573 & 0.3567 & 0.3588 \\
\cline{1-15} \cline{2-15}
\multirow[t]{10}{*}{news} & neutral & this user & 0.6010 & 0.5908 & 0.5823 & 0.5872 & 0.6088 & 0.4593 & 0.4563 & 0.5956 & 0.5773 & 0.5732 & 0.5693 & 0.5806 \\
\cline{2-15}
 & \multirow[t]{3}{*}{gender} & her & 0.5986 & 0.5772 & 0.5671 & 0.5795 & 0.5985 & 0.4357 & 0.4576 & 0.5953 & 0.5736 & 0.5648 & 0.5566 & 0.5664 \\
 &  & him & 0.5980 & 0.5867 & 0.5832 & 0.5832 & 0.6038 & 0.4446 & 0.4511 & 0.5965 & 0.5804 & 0.5628 & 0.5669 & 0.5730 \\
 &  & them & 0.5990 & 0.5842 & 0.5778 & 0.5867 & 0.6055 & 0.4510 & 0.4545 & 0.6002 & 0.5814 & 0.5715 & 0.5647 & 0.5751 \\
\cline{2-15}
 & \multirow[t]{6}{*}{age} 
 & a high school student & 0.5922 & 0.5905 & 0.5924 & 0.5903 & 0.5463 & 0.4670 & 0.5247 & 0.5286 & 0.5042 & 0.5082 & 0.5159 & 0.5535 \\
 & & a college student & 0.5845 & 0.5980 & 0.6005 & 0.5922 & 0.5519 & 0.4629 & 0.5051 & 0.5513 & 0.4817 & 0.4671 & 0.4883 & 0.4987 \\
 &  & a parent of young children & 0.5759 & 0.5623 & 0.5647 & 0.5696 & 0.5211 & 0.4830 & 0.5142 & 0.4991 & 0.4583 & 0.4532 & 0.4600 & 0.4742 \\
 &  & a working professional & 0.5939 & 0.5976 & 0.6040 & 0.6002 & 0.5059 & 0.4722 & 0.5114 & 0.5135 & 0.5216 & 0.5155 & 0.5263 & 0.5510 \\
 &  & a senior citizen & 0.5995 & 0.6029 & 0.6037 & 0.5928 & 0.5686 & 0.4915 & 0.5583 & 0.5675 & 0.5020 & 0.4913 & 0.4941 & 0.5280 \\
 &  & a retired individual & 0.5890 & 0.5949 & 0.5944 & 0.5848 & 0.5463 & 0.4962 & 0.5462 & 0.5541 & 0.4887 & 0.4744 & 0.4827 & 0.5118 \\
\bottomrule
\end{tabular}
}
    
\end{table*}

\subsubsection*{Recommendation Effectiveness}
Full effectiveness scores are available in \Cref{tab:f1_scores_combined}.
Across prompt types and sensitive attribute values, F1 
ranges in $[0.26, 0.44]$ (avg.~$0.39$) for jobs and in $[0.44, 0.61]$ (avg.~$0.55$) for news.
There are no clear trends among different prompts.
The only exception is LLaMa, for which \ac{UR} underperforms all the other prompts for jobs, and \ac{UR}/\ac{BI} underperform the other prompts for news.
Between neutral and sensitive prompts, the F1 difference is minimal for jobs (difference range: $[-0.078, 0.003]$, avg.~$-0.020$) and more pronounced, but still low, for news (difference range: $[-0.119, 0.102]$, avg.~$-0.011$).

\begin{table}[t]
\caption{Recommendation fairness (\ac{SNSR} and \ac{SNSV}) scores for each model and prompt type across sensitive attributes and similarity metrics. The lower the scores, the fairer. The fairest scores across all models and prompts are bolded. The fairest scores per model are underlined.}
\label{tab:snsr_snsv}
\resizebox{\columnwidth}{!}{
\begin{tabular}{lllrrrr|rrrr|rrrr}
\toprule
   & & model & \multicolumn{4}{c|}{Gemma 2 (9B)} & \multicolumn{4}{|c}{LLaMa 3.1 (8B)} & \multicolumn{4}{|c}{Mistral (7B)} \\
   & & prompt & Base & UR & BI & EBI & Base & UR & BI & EBI & Base & UR & BI & EBI \\
  & Attribute & Similarity &  &  &  &  &  &  &  &  &  &  &  &  \\
\midrule
& & & \multicolumn{12}{c}{\textit{Dataset: Jobs}} \\
\midrule
 \multirow[c]{8}{*}{\rotatebox[origin=c]{90}{SNSR}} & \multirow[c]{4}{*}{gender} & Jaccard & \underline{0.061} & 0.167 & 0.173 & 0.180 & 0.077 & 0.158 & 0.121 & \bfseries 0.027 & 0.106 & 0.158 & 0.129 & \underline{0.080} \\
 &  & SERP & \underline{0.045} & 0.122 & 0.120 & 0.151 & 0.042 & 0.123 & 0.118 & \bfseries 0.032 & 0.082 & 0.137 & 0.121 & \underline{0.056} \\
 &  & PRAG & \underline{0.049} & 0.130 & 0.122 & 0.171 & 0.041 & 0.133 & 0.119 & \bfseries 0.028 & 0.083 & 0.158 & 0.147 & \underline{0.062} \\
 &  & BERTScore & \underline{0.048} & 0.125 & 0.132 & 0.151 & 0.040 & 0.108 & 0.127 & \bfseries 0.022 & \underline{0.075} & 0.137 & 0.115 & 0.078 \\
\cline{2-15}
 & \multirow[c]{4}{*}{age} & Jaccard & 0.571 & 0.366 & 0.467 & \underline{0.360} & 0.504 & 0.337 & \bfseries 0.139 & 0.191 & 0.354 & 0.370 & \underline{0.286} & 0.293 \\
  &  & SERP & 0.615 & \underline{0.411} & 0.536 & 0.425 & 0.621 & 0.404 & \bfseries 0.194 & 0.320 & 0.464 & 0.454 & \underline{0.390} & 0.404 \\
  &  & PRAG & 0.624 & \underline{0.429} & 0.537 & 0.435 & 0.605 & 0.423 & \bfseries 0.182 & 0.320 & 0.471 & 0.436 & \underline{0.386} & 0.391 \\
  &  & BERTScore & 0.460 & 0.335 & 0.367 & \underline{0.295} & 0.369 & 0.305 & 0.195 & \bfseries 0.140 & 0.385 & 0.332 & 0.306 & \underline{0.265} \\
\cline{1-15}
  \multirow[c]{8}{*}{\rotatebox[origin=c]{90}{SNSV}} & \multirow[c]{4}{*}{gender} & Jaccard & \underline{0.026} & 0.074 & 0.071 & 0.080 & 0.034 & 0.065 & 0.050 & \bfseries 0.012 & 0.047 & 0.073 & 0.053 & \underline{0.033} \\
 &  & SERP & \underline{0.019} & 0.053 & 0.049 & 0.068 & 0.018 & 0.051 & 0.049 & \bfseries 0.015 & 0.038 & 0.064 & 0.049 & \underline{0.024} \\
 &  & PRAG & \underline{0.020} & 0.057 & 0.050 & 0.076 & 0.017 & 0.055 & 0.049 & \bfseries 0.013 & 0.038 & 0.071 & 0.060 & \underline{0.026} \\
 &  & BERTScore & \underline{0.020} & 0.057 & 0.055 & 0.065 & 0.019 & 0.045 & 0.052 & \bfseries 0.010 & 0.035 & 0.060 & 0.047 & \underline{0.032} \\
\cline{2-15}
   & \multirow[c]{4}{*}{age} & Jaccard & 0.242 & 0.156 & 0.200 & \underline{0.151} & 0.208 & 0.140 & \bfseries 0.054 & 0.070 & 0.146 & 0.147 & \underline{0.114} & 0.119 \\
   &  & SERP & 0.259 & \underline{0.174} & 0.228 & 0.178 & 0.254 & 0.167 & \bfseries 0.078 & 0.117 & 0.190 & 0.175 & \underline{0.151} & 0.161 \\
   &  & PRAG & 0.264 & \underline{0.182} & 0.228 & 0.183 & 0.248 & 0.176 & \bfseries 0.075 & 0.118 & 0.193 & 0.168 & \underline{0.150} & 0.156 \\
   &  & BERTScore & 0.194 & 0.142 & 0.157 & \underline{0.122} & 0.151 & 0.127 & 0.076 & \bfseries 0.057 & 0.149 & 0.129 & 0.114 & \underline{0.105} \\
\midrule
& & & \multicolumn{12}{c}{\textit{Dataset: News}} \\
\midrule
 \multirow[c]{8}{*}{\rotatebox[origin=c]{90}{SNSR}} & \multirow[c]{4}{*}{gender} & Jaccard & 0.223 & 0.265 & \underline{0.134} & 0.195 & 0.138 & 0.125 & \underline{0.068} & 0.091 & \bfseries 0.035 & 0.082 & \bfseries 0.035 & 0.083 \\
   &  & SERP & 0.198 & 0.307 & \underline{0.140} & 0.222 & 0.127 & 0.135 & \underline{0.065} & 0.106 & 0.070 & 0.117 & \bfseries 0.057 & 0.111 \\
   &  & PRAG & 0.192 & 0.319 & \underline{0.143} & 0.228 & 0.124 & 0.136 & \underline{0.073} & 0.114 & \bfseries 0.063 & 0.126 & \bfseries 0.063 & 0.118 \\
   &  & BERTScore & 0.152 & 0.281 & \underline{0.129} & 0.196 & 0.113 & 0.105 & \bfseries 0.081 & 0.105 & \underline{0.082} & 0.159 & 0.086 & 0.161 \\
\cline{2-15}
   & \multirow[c]{4}{*}{age} & Jaccard & \bfseries 0.000 & 0.006 & 0.002 & 0.007 & \underline{0.005} & 0.011 & 0.010 & 0.006 & \underline{0.002} & 0.006 & 0.003 & 0.005 \\
   &  & SERP & \bfseries 0.001 & 0.008 & 0.002 & 0.006 & \underline{0.008} & 0.018 & 0.022 & 0.014 & \underline{0.004} & 0.009 & 0.006 & 0.008 \\
   &  & PRAG & \bfseries 0.001 & 0.009 & 0.004 & 0.004 & \underline{0.008} & 0.010 & 0.023 & 0.011 & \underline{0.004} & 0.008 & 0.007 & 0.008 \\
   &  & BERTScore & \bfseries 0.010 & 0.036 & 0.018 & 0.016 & 0.036 & \underline{0.023} & 0.097 & 0.068 & 0.063 & 0.062 & \underline{0.034} & 0.040 \\
\cline{1-15}
  \multirow[c]{8}{*}{\rotatebox[origin=c]{90}{SNSV}} & \multirow[c]{4}{*}{gender} & Jaccard & 0.092 & 0.116 & \underline{0.060} & 0.090 & 0.062 & 0.051 & \underline{0.028} & 0.039 & 0.016 & 0.034 & \bfseries 0.015 & 0.039 \\
   &  & SERP & 0.081 & 0.131 & \underline{0.061} & 0.103 & 0.058 & 0.056 & \underline{0.027} & 0.045 & 0.029 & 0.048 & \bfseries 0.024 & 0.052 \\
   &  & PRAG & 0.079 & 0.135 & \underline{0.062} & 0.106 & 0.057 & 0.056 & \underline{0.030} & 0.048 & \bfseries 0.026 & 0.051 & \bfseries 0.026 & 0.055 \\
   &  & BERTScore & 0.063 & 0.121 & \underline{0.054} & 0.092 & 0.049 & 0.043 & \bfseries 0.034 & 0.044 & \bfseries 0.034 & 0.066 & 0.037 & 0.070 \\
\cline{2-15}
   & \multirow[c]{4}{*}{age} & Jaccard & \bfseries 0.000 & 0.002 & 0.001 & 0.002 & \underline{0.002} & 0.004 & 0.004 & \underline{0.002} & \underline{0.001} & 0.002 & \underline{0.001} & 0.002 \\
  &  & SERP & \bfseries 0.000 & 0.003 & 0.001 & 0.002 & \underline{0.003} & 0.006 & 0.008 & 0.004 & \underline{0.001} & 0.003 & 0.002 & 0.003 \\
  &  & PRAG & \bfseries 0.000 & 0.003 & 0.001 & 0.001 & \underline{0.002} & 0.003 & 0.009 & 0.003 & \underline{0.001} & 0.003 & 0.002 & 0.003 \\
  &  & BERTScore & \bfseries 0.003 & 0.012 & 0.006 & 0.005 & 0.012 & \underline{0.008} & 0.034 & 0.023 & 0.022 & 0.022 & \underline{0.014} & 0.016 \\
\bottomrule
\end{tabular}
}
\end{table}
\begin{figure}[tb]
    \centering
    \includegraphics[width=\linewidth,trim=0 0.25cm 0 0, clip=True]{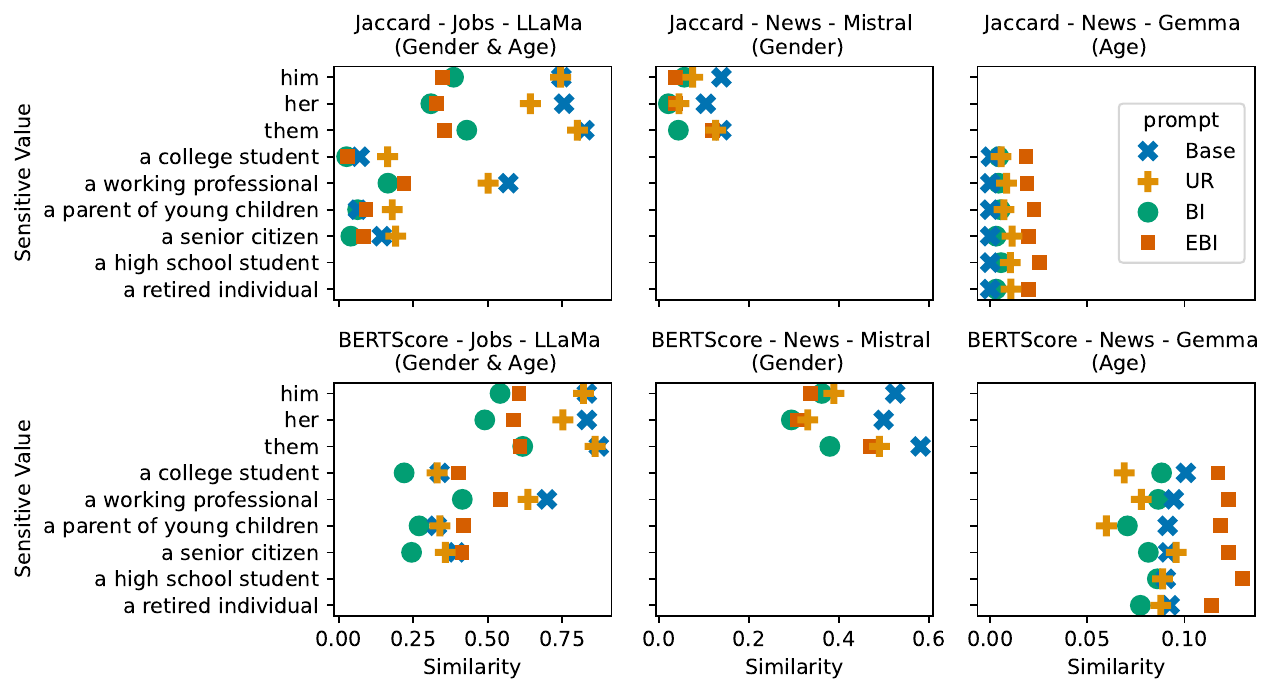}
    \caption{Recommendation similarity of neutral vs.~sensitive variants, with Jaccard (top) and BERTScore (bottom) for the fairest LLMs.
    }
    \label{fig:sim_sensitive_to_neutral}
    \includegraphics[width=\linewidth, trim=0 0.22cm 0 0, clip=True]{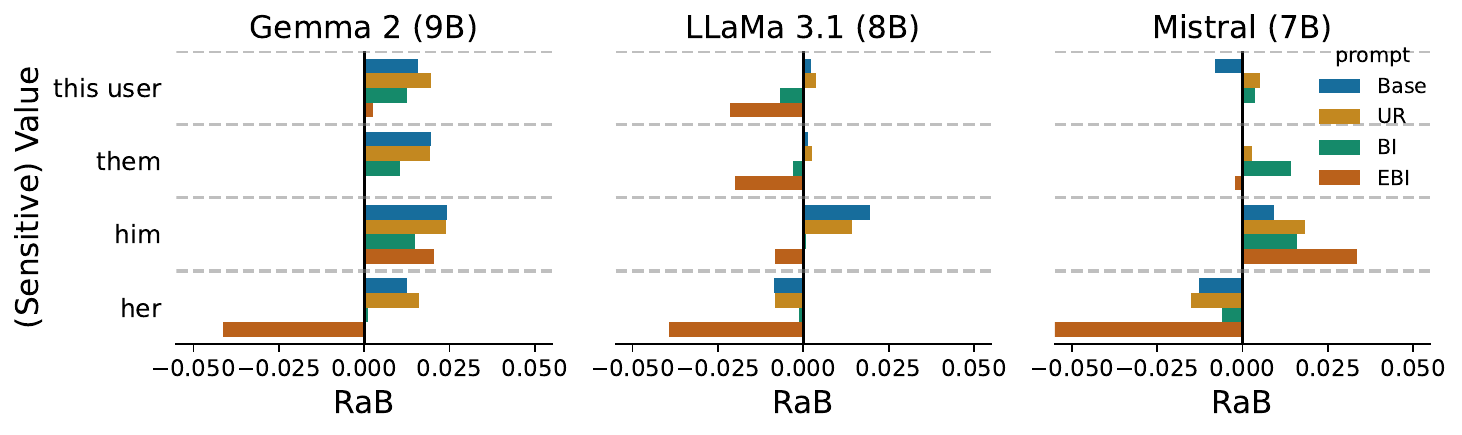}
    \caption{Gender bias (RaB) of news recommendation, with gender as a sensitive attribute. RaB>0 means the output has more male- than female-gendered words. Vice versa for < 0.}
    \label{fig:gender_bias}
\end{figure}

\subsubsection*{Fairness across groups}
\Cref{tab:snsr_snsv} reports fairness with \ac{SNSV} and \ac{SNSR}.
The best \ac{LLM}-prompt combination is LLaMa with \ac{BI} and \ac{EBI} on jobs. Specifically, LLaMA with \ac{BI} improves SNSV (Jaccard) of Base by up to 74.0\% ($0.208\rightarrow0.054$) on jobs with age attributes.
On news, the trend is less clear; Gemma with Base and Mistral with Base/\ac{BI} are the fairest (up to $46\%$ SNSR BERTScore improvement, $0.063\rightarrow0.034$).
Fairness scores for news with age values tend to be close to $0$ because the average similarity between neutral and sensitive prompt outputs is also close to $0$ (see~\Cref{fig:sim_sensitive_to_neutral}), and not because \acp{LLMRec} are exceptionally fair (i.e., having high  similarity between neutral and sensitive prompt outputs).
This shows that age values have a large impact on the \acp{LLMRec}, which generate very different outputs when prompted with and without age.
Finally, \acp{LLMRec} tend to be fairer for gender than age in jobs, but the opposite happens for news.
This suggests that gender values affect news recommendations more, while age values affect job recommendations more.

In general, bias-aware prompts perform better than the baseline prompt, with a few exceptions.
For Gemma on jobs, \ac{UR} and \ac{EBI} perform better for age, while Base is better for gender.
On news, \ac{BI} is better for gender, while Base is better for age, even if the scores are overall very small.
For LLaMa and Mistral on jobs, \ac{BI} and \ac{EBI} are the best performing prompts with gender and age.
On news, \ac{BI} performs the best for gender, but for age, Base is better, again with only marginal differences to the bias-aware prompts.
Finally, prompts which include instructions to avoid bias (\ac{BI}, \ac{EBI}) perform better than the role prompt (\ac{UR}).
This differs from~\citep{furniturewala2024thinking}, where role prompts lead to fairer results than instruction prompts.

\subsubsection*{Similarity to neutral recommendation} 
\ac{SNSV} and \ac{SNSR} only quantify how much the similarity between the neutral and sensitive prompts varies across sensitive values; they do not account for how similar the neutral and sensitive prompt outputs are, i.e., the average similarity score.
\Cref{fig:sim_sensitive_to_neutral} shows Jaccard and BERTScore between the recommendations generated with the neutral and sensitive prompt variants for the best combinations in~\Cref{tab:snsr_snsv}.
Results for \ac{SERP} and \ac{PRAG} are aligned with Jaccard.

Overall, the similarity between the neutral and sensitive prompts tend to be higher for gender than age, meaning that the output of \acp{LLMRec} is less affected when prompted with gender than age values.
We hypothesize that this is due to the extensive work done on gender debiasing \acp{LLM} \cite{Bartl2025GenderBiasSurvey, stanczak2021surveygenderbiasnatural}.
Among prompt types, Base and \ac{UR} exhibit the highest similarity between the neutral and sensitive variants for gender, while \ac{EBI} has the highest similarity for age.
However, since fairness is computed as the min-max difference (\ac{SNSV}) and standard deviation (\ac{SNSR}) of similarity across sensitive values, these prompts do not necessarily lead to the fairest \acp{LLMRec}.

Furthermore, similarity scores tend to be higher and spread on a wider range when computed with BERTScore because Jaccard only accounts for overlapping words, while BERTScore considers semantic similarity.
Therefore, BERTScore might be a viable alternative to measure similarity of \acp{LLMRec} outputs, which likely differ in wording but retain comparable meanings.

\subsubsection*{Overadjustment} 
To measure the extent of over-adjusted recommendations for different genders, we compute Ranking Bias (RaB) metric \cite{Rekabsaz2020RaB}. RaB considers the log difference of the number of male- and female-gendered words in an item title, averaged across all recommended items; RaB>0 means that there are more male- than female-gendered words, while RaB<0 means the opposite. 
The list of gendered words is obtained from \cite{Rekabsaz2020RaB}. As job titles usually do not contain gendered words, we only evaluate the news dataset. 
\Cref{fig:gender_bias} presents the mean RaB of all users per \ac{LLM} and prompt type. 

\Cref{fig:gender_bias} shows that Gemma tends to return more male-gendered words than LLaMa and Mistral across all prompts.
The \ac{BI} prompt is the most neutral across gender values and \acp{LLM}, while \ac{EBI} is the most sensitive, likely because it explicitly refers to the attribute type.
When the \ac{EBI} prompt includes `her', there is a disproportionate use of female-gendered words. Manually analyzing the responses, we find that `her' prompts output more women-related news (e.g., women’s achievements, women’s sports, see \Cref{fig:teaser}), while `him'/`them' prompt outputs are largely the same as before.

\section{Conclusions and Future Work}

Experiments with $12$ combinations of \acp{LLM} and prompts on $2$ high-stakes \ac{RS} datasets show that fairness improves when \acp{LLM} are instructed to avoid bias. However, a closer inspection of the \ac{LLMRec} outputs reveals that in some cases recommendations are over-adjusted to a specific demographic group, e.g., women empowerment news. In addition, we find that BERTScore distinguishes recommendations generated with and without sensitive attributes better than metrics based on exact matches; we encourage using BERTScore in \acp{LLMRec} fairness evaluation. Future work can extend BERTScore to consider the semantic similarity of items at different rank positions, for a more lenient evaluation. Overall, bias-aware prompting strategies have promising results, and future work should compare them to other bias mitigation strategies as well as investigate strategies to avoid over-adjustment of \acp{LLMRec}.

\begin{acks}
The work is supported by the Algorithms, Data, and Democracy project (ADD-project), funded by Villum Foundation and Velux Foundation. 
We thank Shivam Adarsh and Pietro Tropeano for their helpful feedback on the manuscript.
\end{acks}

\bibliographystyle{ACM-Reference-Format}
\bibliography{sample-base}

\end{document}